\documentclass[aps,prl,twocolumn,superscriptaddress,showpacs,floatfix,10pt]{revtex4-1}

\usepackage{amsmath}
\usepackage{graphicx}
\usepackage{color}
\usepackage{epstopdf}
\usepackage{natbib}
\usepackage{xfrac}
\usepackage{wasysym}
\usepackage{multirow}

\usepackage[margin=10pt,format=hang,justification=raggedright,singlelinecheck=false]{subfig}%[justification=raggedright,singlelinecheck=false]

\usepackage{xcolor,colortbl}
\definecolor{Gray}{gray}{0.85}
\definecolor{lblue}{rgb}{0.8,0.8,1}  %{blue}{0.5}
\definecolor{blue}{rgb}{0.6,0.6,0.9}  %{blue}{0.5}

\newcommand{\cbp}{Ce$_3$Bi$_4$Pt$_3$}

\newcommand{\eref}[1]{Eq.~(\ref{#1})}
\newcommand{\fref}[1]{Fig.~\ref{#1}}
\newcommand{\Fref}[1]{Fig.~\ref{#1}}

\newcommand{\vek}[1]{ \hbox{\textbf #1}}
\newcommand{\svek}{\mathbf}

\newcommand{\etal}{{\it et al.}}

\renewcommand{\Im}{\hbox{Im}}
\renewcommand{\Re}{\hbox{Re}}

\newcommand{\out}[1]{}

\begin{document}
\title{Realistic many-body theory of Kondo insulators:\\ Renormalizations and fluctuations in \cbp}

\author{Jan M.~Tomczak}
\email{tomczak.jm@gmail.com}
\affiliation{Institute for Solid State Physics, TU Wien,  Vienna, Austria}

\date{\today}

\begin{abstract}
Our theoretical understanding of heavy-fermion compounds mainly derives from iconic models, such as those due to Kondo or Anderson. 
While providing invaluable qualitative insight, detailed comparisons to experiments are
encumbered by the materials' complexity, 
including the spin-orbit coupling, crystal fields, and ligand 
hybridizations. 
Here, we study the paradigmatic Kondo insulator \cbp\ with a first principles dynamical mean-field method that includes these complications.
We find that salient signatures of many-body effects in this
material---large effective masses, the insulator-to-metal crossover, the concomitant emergence of Curie-Weiss behaviour and notable transfers of optical spectral weight---are captured {\it quantitatively}. 
With this validation, we elucidate the fabric of the many-body state. 
In particular, we extent the phenomenology of the Kondo crossover to time-scales of fluctuations:
We evidence that spin and charge degrees of freedom each realize two regimes in which fluctuations adhere to vastly different decay laws.
We find these regimes to be separated by a {\it common} temperature $T^{max}_\chi$, linked to the onset of Kondo screening.
Interestingly, below (above) $T^{max}_\chi$, valence fluctuations become faster (slower) than the dynamical screening of the local moments.
Overall, however, spin and charge fluctuations occur on comparable time-scales of $\mathcal{O}(0.5-12\hbox{ fs})$,
placing them on the brink of detection for modern time-resolved probes.
\end{abstract}

%\keywords{}

\maketitle

%%%%%%%%%%%%%%%%%%%%%%%%%%%%%%%%%%%%%%%%%%%%%%%%%%%%%%%%%%%%%%%%%%%%%%%%%%%%%%%%%%%%%%%%%%%%%%%%%%%%%%%%%%%%%%%%%%%%%%%%%%%%%%%%
\paragraph{Introduction.---}%
%%%%%%%%%%%%%%%%%%%%%%%%%%%%%%%%%%%%%%%%%%%%%%%%%%%%%%%%%%%%%%%%%%%%%%%%%%%%%%%%%%%%%%%%%%%%%%%%%%%%%%%%%%%%%%%%%%%%%%%%%%%%%%%%
%
%
Heavy-fermion materials\cite{Wirth2016,Colemann2007,PSSB:PSSB201300005} harbour
a cornucopia of phenomena: the Kondo effect, spin-ordered phases, superconductivity, quantum criticality, and, potentially, topological effects. 
For a subclass of these system, coined Kondo insulators\cite{ki,Riseborough2000,NGCS}, the heavy  mobile charges released via Kondo screening completely fill the valence bands,
thus opening a charge and spin gap.
Qualitatively, this behaviour can be understood, e.g., within the periodic Anderson model\cite{PhysRevLett.70.1670,PhysRevB.54.8452}. 
However, oftentimes the physics in real materials is strongly influenced by ingredients not contained
in reductionist models. Given the delicate competition of a multitude of energy scales in correlated materials, leaving out complexity may 
substantially bias the analysis%
\footnote{E.g., pioneering model studies\cite{PhysRevB.51.17439,PhysRevB.78.033109,sentef:155116} suggested the
$d$-electron ``flyweight''\cite{Colemann2007} Kondo insulator FeSi\cite{NGCS} to be controlled by Hubbard physics.
Then, realistic calculations\cite{jmt_fesi,jmt_hvar} 
divulged the Hund's rule coupling---a genuine multi-orbital effect\cite{annurev-conmatphys-020911-125045}---to drive electronic correlations,
explaining\cite{NGCS} why---instead of {\it local} (Curie-Weiss) moments\cite{PhysRevB.51.17439,PhysRevB.78.033109}---neutron experiments\cite{PhysRevLett.59.351,PhysRevB.38.6954} see a susceptibility enhanced at ferromagnetic wavevectors.}.
Indeed, electronic structure details have been identified as crucial for understanding heavy-fermion insulators\cite{Doniach1994450,Shick2015,Wissgott2016,PhysRevX.7.011027}.
%
%\footnote{A case in point is the Kondo insulator to semi-metal crossover in Ce$_3$Bi$_4$(Pt$_{1-x}$Pd$_x$)$_3$\cite{PhysRevLett.118.246601}.
%As the series is iso-volume, an $x$-independent Kondo coupling was advocated, implying the crossover to be driven by the spin-orbit coupling. 
%Subsequent calculations\cite{NGCS,jmt_radialKI} instead showed that the Kondo coupling varies
%notably, owing to the different radial extent of the $4d$ ($5d$) orbitals of Pd (Pt).
%}.
%
The most formidable challenge is to quantitatively capture the signatures of Kondo-singlet formation in
spectral, optical and magnetic observables over a large temperature range\cite{Colemann2007}.
Important insights have been gained under simplifying circumstances---low temperatures\cite{Shick2015,PhysRevX.7.011027}, high temperatures\cite{Wissgott2016}, weak hybridizations\cite{jmt_cesf}, or
strong coupling\cite{2013arXiv1312.6637D}---yet an unbiased description of coherence effects in 4$f$-systems still heavily relies on model calculations\cite{Coleman_Juelich,Min2017}\footnote{see, however, the recent Ref.\ \onlinecite{Goremychkin186}}. 
Here, we spearhead realistic simulations that meet all of the above challenges for the archetypal Kondo insulator \cbp.
After analysing signatures of Kondo screening in diverse observables, we
characterize the time-dependence of spin and charge relaxation processes.
By establishing the magnitude of relevant time-scales, 
we provide essential guidelines for future  applications of time-resolved probes\cite{Hentschel2001,Goulielmakis2010,Buzzi2018} to Kondo insulators.
% The latter 

\begin{figure*}[!th]
\hspace{-0.25cm}
 {\includegraphics[width=0.345\textwidth]{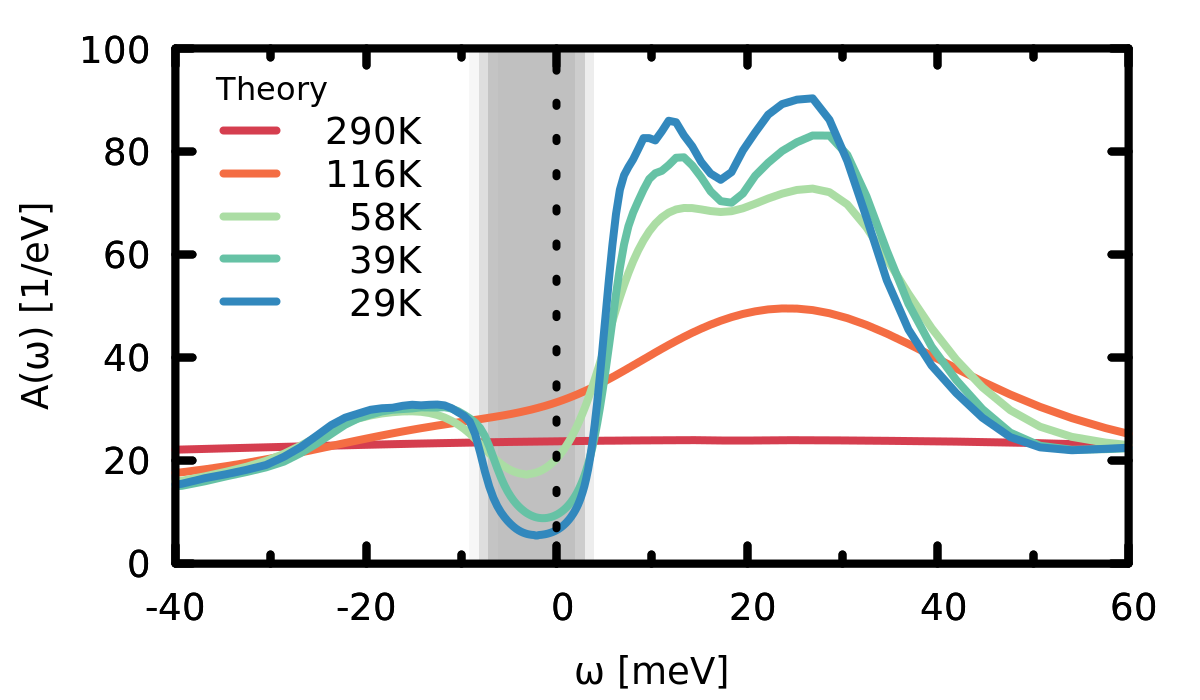}}
\hspace{-0.42cm}
{\includegraphics[width=0.345\textwidth]{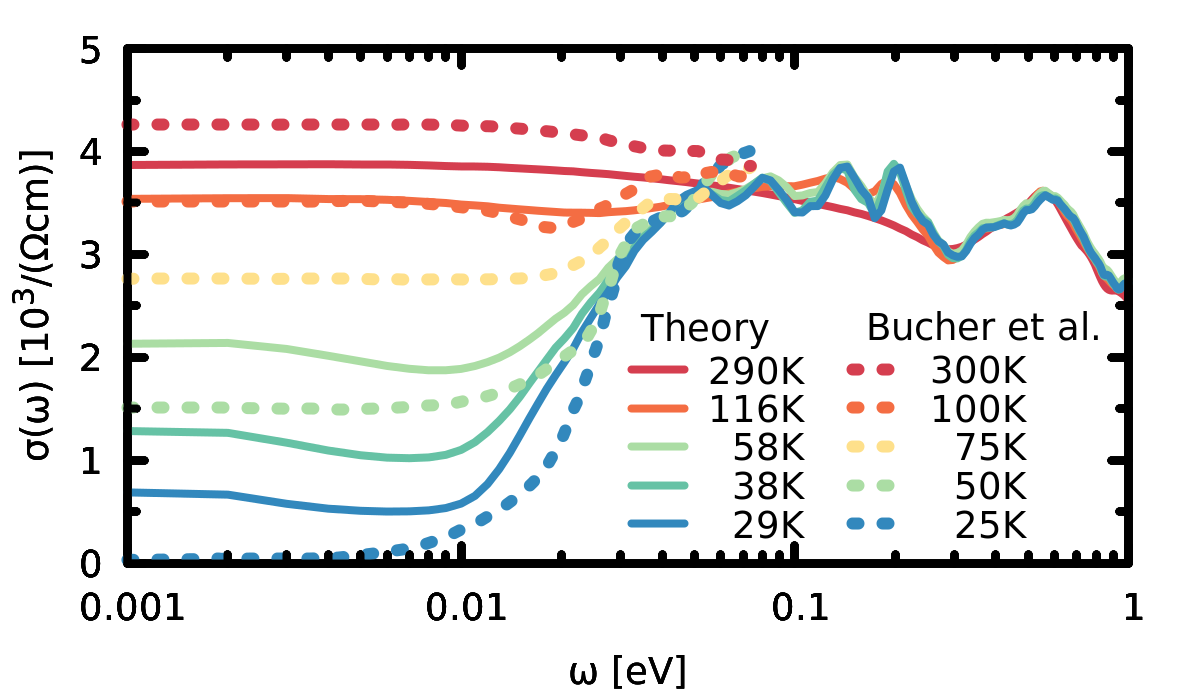}}
\hspace{-0.42cm}
{\includegraphics[width=0.345\textwidth]{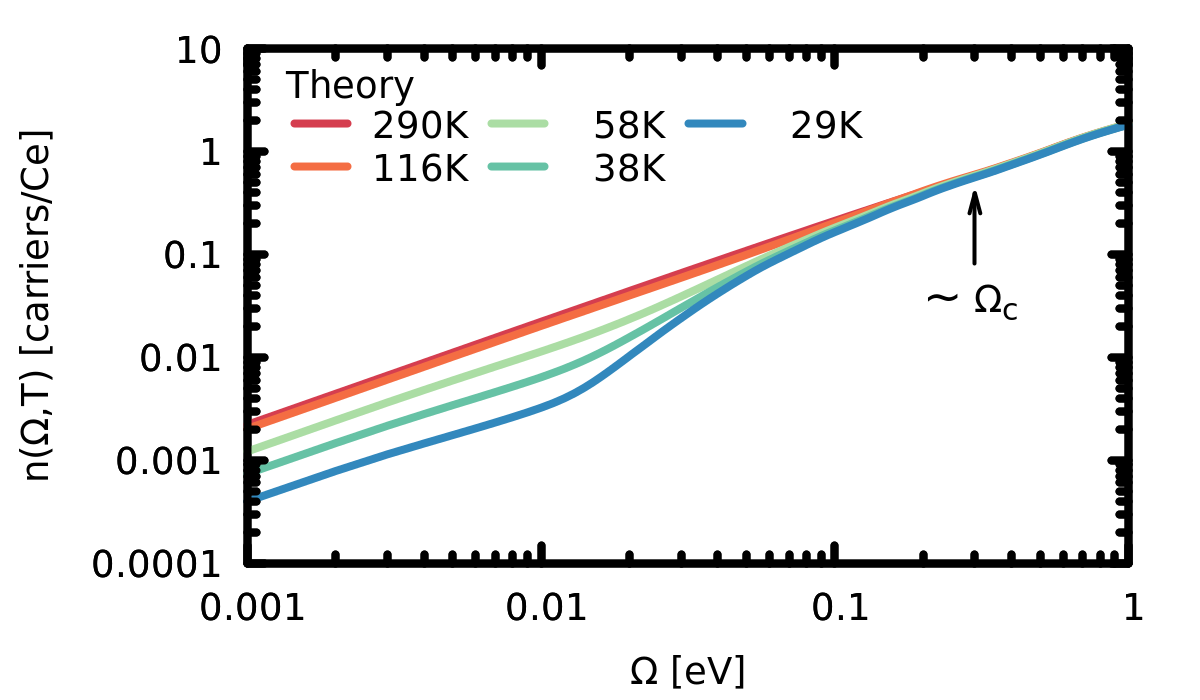}}
\hspace{-0.25cm}
\caption{(color online). Spectral and optical properties: 
(a) local spectral function $A(\omega)$, (b) optical conductivity $\sigma(\omega)$ (experiment from Ref.~\onlinecite{PhysRevLett.72.522}), (c) optical weight $n(\Omega,T)=\frac{2m}{\hbar\pi e^2}\int^\Omega_0\sigma(\omega,T)d\omega$ is recovered only for $\Omega_c\gtrsim 300$meV.
} \label{spec}
\end{figure*}

\paragraph{Prelude.---}%
The cubic intermetallic compound \cbp\ is a classic  Kondo insulator\cite{PhysRevB.42.6842,Riseborough2000,NGCS}:
%Experimental observables, including 
Photoemission\cite{Takeda1999721} and optical\cite{PhysRevLett.72.522} spectra, as well as magnetic susceptibilities\cite{PhysRevB.42.6842,PhysRevLett.118.246601} 
are qualitatively consistent with the Kondo scenario\cite{DONIACH1977231}. 
Moreover, resistivity\cite{BOEBINGER1995227} and specific heat\cite{Jaime2000}
measurements in high magnetic fields evidenced a collapse of the charge and spin gap---congruent with
%expectations of 
a field-induced coming asunder of Kondo singlets.
Interest in \cbp\ surged by the recent proposal\cite{Chang2017}
that it may harbour topological effects 
associated with its non-symmorphic crystal symmetry\cite{PhysRevB.91.155120}.
Indeed, topological Kondo insulators\cite{Dzero2016} 
provide a unique playground to study the interplay of electronic correlations and topology\cite{0034-4885-79-9-094504}%
\footnote{in
stark contrary to Mott-Hubbard insulators in which interaction effects can severely tamper with topological protections
based on single particle symmetries\cite{DiSante_cubio}.}.

\paragraph{Method.---}%
Realistic many-body calculations were performed in the framework of density functional theory plus dynamical-mean field theory (DFT+DMFT)\cite{bible,doi:10.1080/00018730701619647},
employing the package of Haule \etal\ \cite{PhysRevB.81.195107,PhysRevB.75.155113}---that combines DFT from Wien2k\cite{wien2k}
with a continuous-time quantum monte-carlo\cite{RevModPhys.83.349} solver %via a projection formalism 
and includes charge self-consistency---which has previously also been applied to
 elemental Ce\cite{PhysRevB.75.155113,PhysRevB.89.125113} and Ce-compounds\cite{PhysRevB.81.195107,PhysRevLett.108.016402,Goremychkin186}. We used the PBE functional and rotationally invariant interactions parametrized by a Hubbard $U=5.5$eV and Hund's $J=0.68$eV similar to values used before\cite{PhysRevLett.87.276404,amadon:066402,PhysRevB.81.195107,jmt_cesf,Goremychkin186}.
Analytical continuation was done as described in Ref.~\cite{PhysRevB.81.195107}; optical spectra were computed as in Ref.~\cite{jmt_fesi}.

\begin{figure*}[!t]
\hspace{-0.25cm}
 \includegraphics[width=0.345\textwidth,clip=true,trim=60 60 60 60]{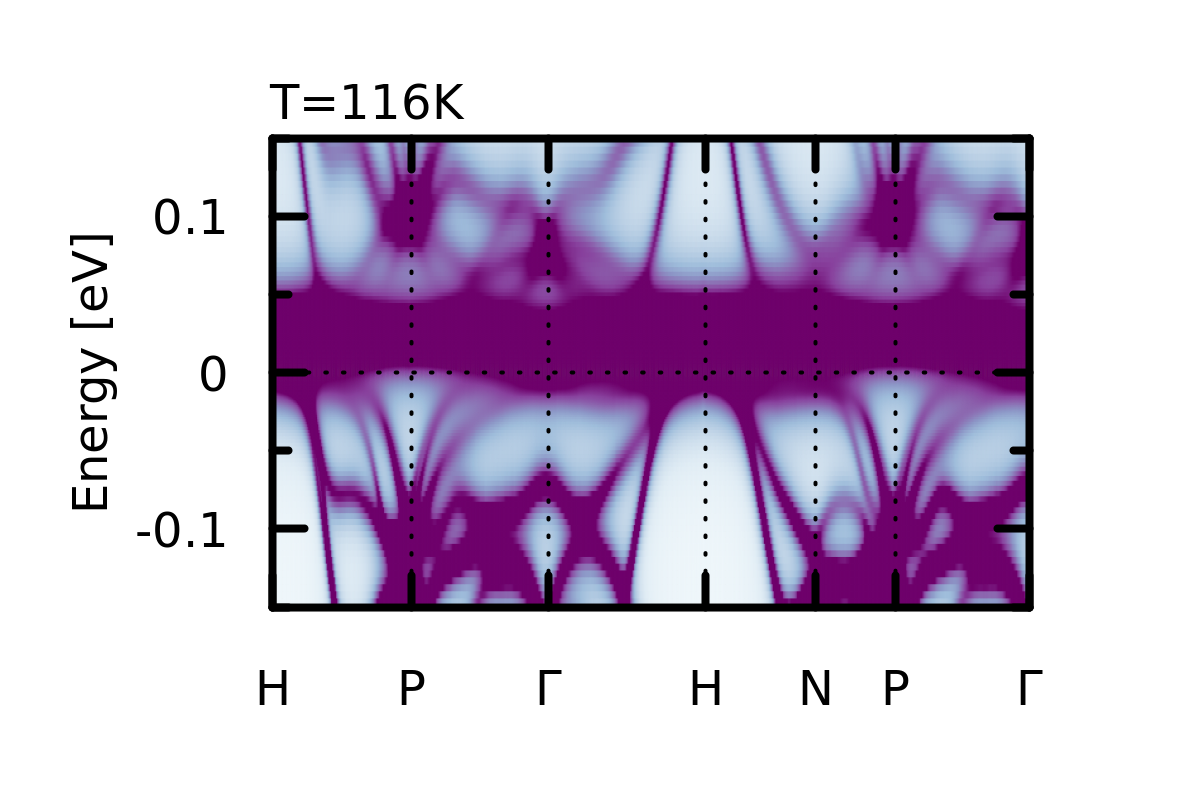}
\hspace{-0.42cm}
 \includegraphics[width=0.345\textwidth,clip=true,trim=60 60 60 60]{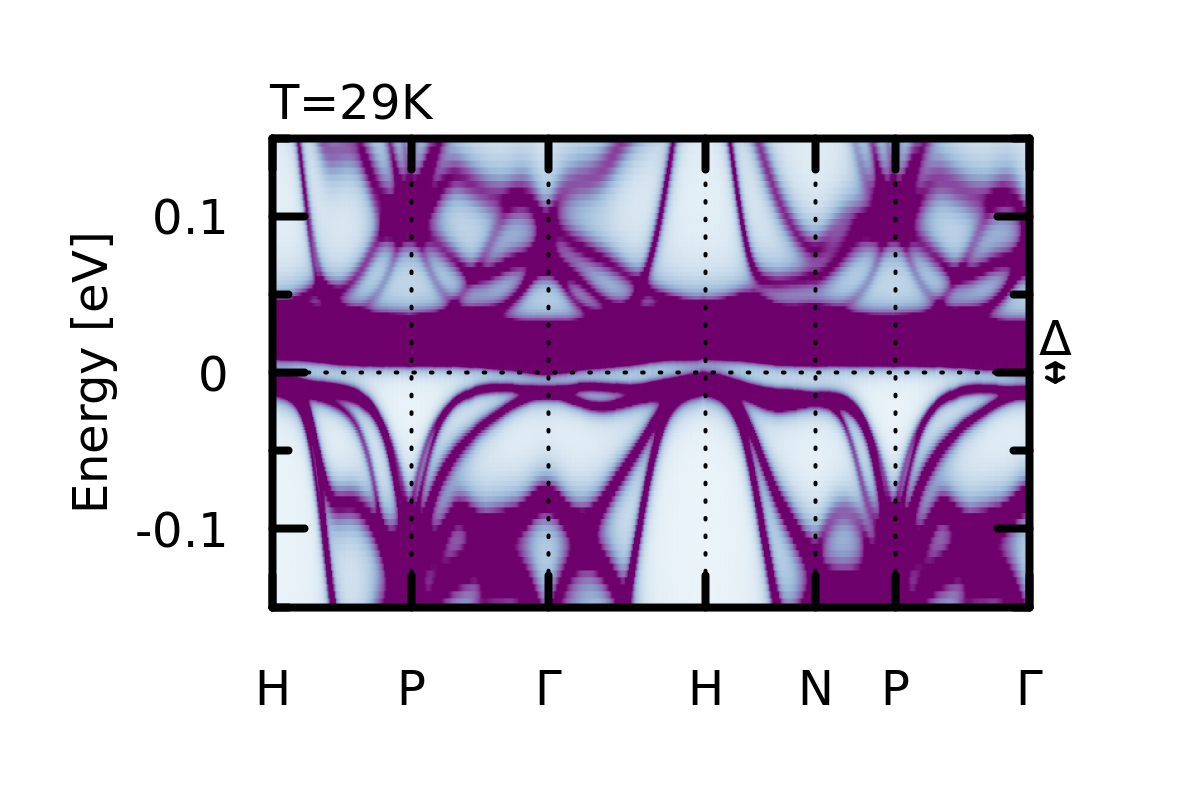}
\hspace{-0.42cm}
\includegraphics[width=0.345\textwidth,clip=true,trim=5 0 30 30]{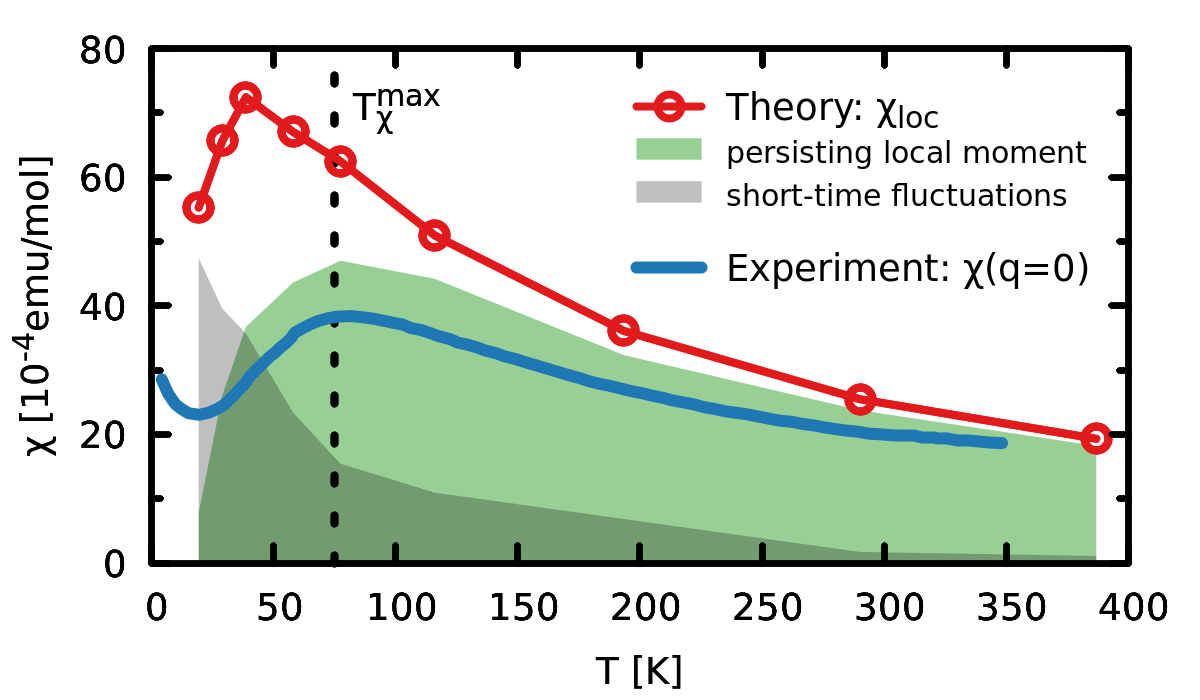}
\hspace{-0.25cm}
 \caption{(color online). Simulated momentum-resolved spectra for (a) $T=116$K (metallic), (b) $T=29$K (insulating with gap $\Delta$). (c) Theoretical local magnetic susceptibility $\chi_{loc}$ and experimental uniform susceptibility $\chi(\svek{q}=0)$\cite{PhysRevB.42.6842}. The former is decomposed via \eref{chidecomp} into fluctuations at short-time scales 
and the local moment that persists for long times (see text for details). The maximum $T^{max}_\chi$ is a proxy for the onset of Kondo screening.
}
 \label{arpessus}
\end{figure*}

\paragraph{Many-body theory: observables.---}%
We first validate the approach by computing experimental observables.
\Fref{spec}(a) displays the simulated local spectral function $A(\omega)$ of \cbp: 
Cooling down from room temperature, a largely featureless metal evolves into a narrow-gap semiconductor with 
a gap $\Delta\approx 10$meV---in congruence with
photoemission spectroscopy (see e.g., Ref. \onlinecite{Takeda1999721}). 
Momentum-resolved spectra reveal more details:
Above 100K, see \fref{arpessus}(a), a wide band of incoherent excitations---incompatible with the notion of a Fermi surface---lingers at the Fermi level. 
Upon cooling, see \fref{arpessus}(b), features outside $\pm 50$meV
sharpen, yet remain roughly at constant positions. Closer to the Fermi level, the effects are much larger: 
Characteristic of Kondo insulators, a charge gap $\Delta$ emerges below a temperature $T^*\sim 50$K (cf.\ \fref{spec}(a)) that is much smaller than the size of the gap, 
$\Delta/k_B\gtrsim 120$K,
which itself is radically smaller than the local Coulomb repulsion $U$.

That our %many-body 
simulations 
capture the metal-insulator crossover in \cbp\ on a {\it quantitative} level is shown in \fref{spec}(b)
for the optical conductivity $\sigma(\omega)$: Direct comparison to experiments of Bucher \etal \cite{PhysRevLett.72.522} reveals that, both, the absolute magnitude of the response
and the energy scale of the (optical) gap that emerges at low temperatures are well reproduced.
Note that the insulator-to-metal crossover is not driven by a displacement of the gap edges. Indeed,
peak positions in $\sigma(\omega)$ do not move significantly with temperature.
Instead, a featureless low-energy response---indicative of incoherent charge carriers---builds up with increasing $T$.
We can analyze the evolution of the optical spectrum through the $f$-sum rule\cite{RevModPhys.71.687}:
The integral, % over the optical conductivity, 
$n(\Omega,T)=\frac{2m}{\hbar\pi e^2}\int^\Omega_0\sigma(\omega,T)d\omega$, 
evaluates---for $\Omega\rightarrow \infty$---to a temperature-independent constant: the total density of carriers participating in optical absorption.
The characteristic frequency $\Omega_c$ above which $n(\Omega,T)$ becomes independent of $T$ thus sets the energy scale over which spectral
weight transfers occur as the system passes through the metal-insulator crossover. 
Experimentally $\Omega_c\sim250$meV\cite{FISK1995798}, corresponding to more than 
five times %the size of 
the optical gap\cite{PhysRevLett.72.522}, indicative of strong correlation effects\cite{PhysRevB.54.8452}.
In our calculation, we capture this energy scale quantitatively, see \fref{spec}(c): Indeed, the loss of low-energy spectral weight upon cooling
is not recovered until above at least $\Omega_c\gtrsim300$meV.
These findings strongly support the hypothesis that states from large parts of the conduction electron bandwidth participate in the formation of Kondo singlets\cite{Colemann2007}.

\paragraph{Many-body theory: microscopic insights.---}%
With the above validation of the computational approach, we now analyse our results in detail. 
We begin with the self-energy that contains pertinent information about many-body renormalizations.
We find the Ce-$4f$-derived spectral weight at the Fermi level to be dominated by the $J=5/2$ components, and 
the corresponding self-energies $\Sigma$ to exhibit characteristics of a Fermi liquid: The real parts are linear at low energy,
giving rise to a mass enhancement of $m^*/m=1/Z=1-\partial_\omega\Re\Sigma(\omega=0)\approx 10$ for all $J=5/2$ components. 
These masses are in line with the shrinking of the charge gap by one order of magnitude with respect to the DFT band-gap ($\Delta_{DFT}\approx130$meV\cite{NGCS,doi:10.1143/JPSJ.62.2103}),
and are only weakly temperature-dependent.
The scattering rates (inverse lifetimes) encoded in the imaginary part of the self-energy 
evolve quadratically with temperature\cite{supps}:
$-\Im\Sigma(\omega=0)=aT^2$ for all $J=5/2$ components, with a prefactor $a=7.8\cdot 10^{-3}$meV/K$^2$ large enough to result in a broadening $-Z\Im\Sigma(0)$ of one-particle excitations in excess of $\Delta\approx10$meV at $T=116$K, cf.\ \fref{arpessus} (a).
These insights quantify the above claim that \cbp\ metallizes through the emergence of incoherent ($\Im\Sigma$-caused)
weight at low-energies, while the fundamental gap (controlled by $\Re\Sigma$) 
remains roughly unchanged.

We continue the analysis by looking at crucial ingredients to the here employed dynamical mean-field theory (DMFT)\cite{bible}: it  maps the periodic solid onto an effective atom  that can exchange particles with a bath, realizing a single-site Anderson impurity model.
The effective atom is characterized by a local Hamiltonian $H^{atom}=H^{loc}+H^{int}$ (consisting of a one-particle and interacting part, respectively)
and the coupling to its environment is described by 
an auxiliary hybridization function, 
$\Delta=\omega+\mu-H^{loc}_{ff}-\Sigma_{ff}-\left(G_{loc}^{-1}\right)_{ff}$. Here $H^{loc}$, $\Sigma$ and $G_{loc}$ are the one-particle part of the atom's Hamiltonian, the impurity self-energy and the locally projected Green's function, respectively;  subscripts $f$ denote a restriction onto a subspace of correlated orbitals (here the Ce-$4f$); $\mu$ is the chemical potential.
Already on the level of band-theory ($\Sigma=0$, $H^{loc}$ from DFT, $H^{int}=0$), $\Delta(\omega)$ can be used to investigate {\it trends} in families of heavy-fermion materials\cite{PhysRevMaterials.1.033802,jmt_radialKI}, e.g., regarding the
degree of itineracy of $4f$-states, the unit-cell volume or the Kondo temperature. 
Within DMFT, $\Delta(\omega)$ 
is promoted to the
Weiss mean-field, which is determined self-consistently through the requirement
\begin{eqnarray}
&&\left[\sum_{\svek{k}}\left[\omega+\mu-H(\svek{k})-\Sigma_{ff}(\omega) \right]^{-1}\right]_{ff}\nonumber\\
&&\qquad\qquad=\left[\omega+\mu-H_{ff}^{loc}-\Delta(\omega)-\Sigma_{ff}(\omega)  \right]^{-1}
\label{SCC}
\end{eqnarray}
stating that the local projection of the solid's Green's function in the subspace of correlated orbitals (l.h.s.\ of \eref{SCC}) equals
the Green's function of the impurity (r.h.s.)---when approximating the solid's self-energy $\Sigma(\svek{k},\omega)$ by the impurity $\Sigma_{ff}(\omega)$ one.
 
\begin{figure*}[!t]
 \hspace{-0.25cm}
\includegraphics[width=0.345\textwidth]{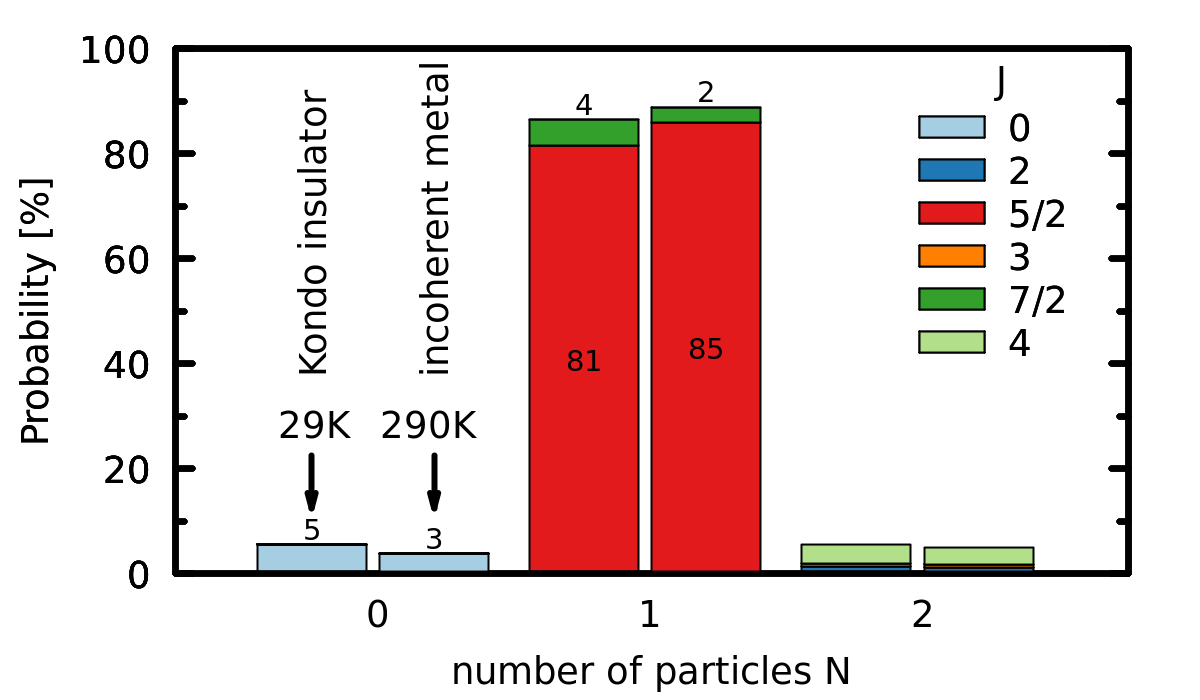} 
\hspace{-0.4cm}
 \includegraphics[width=0.345\textwidth]{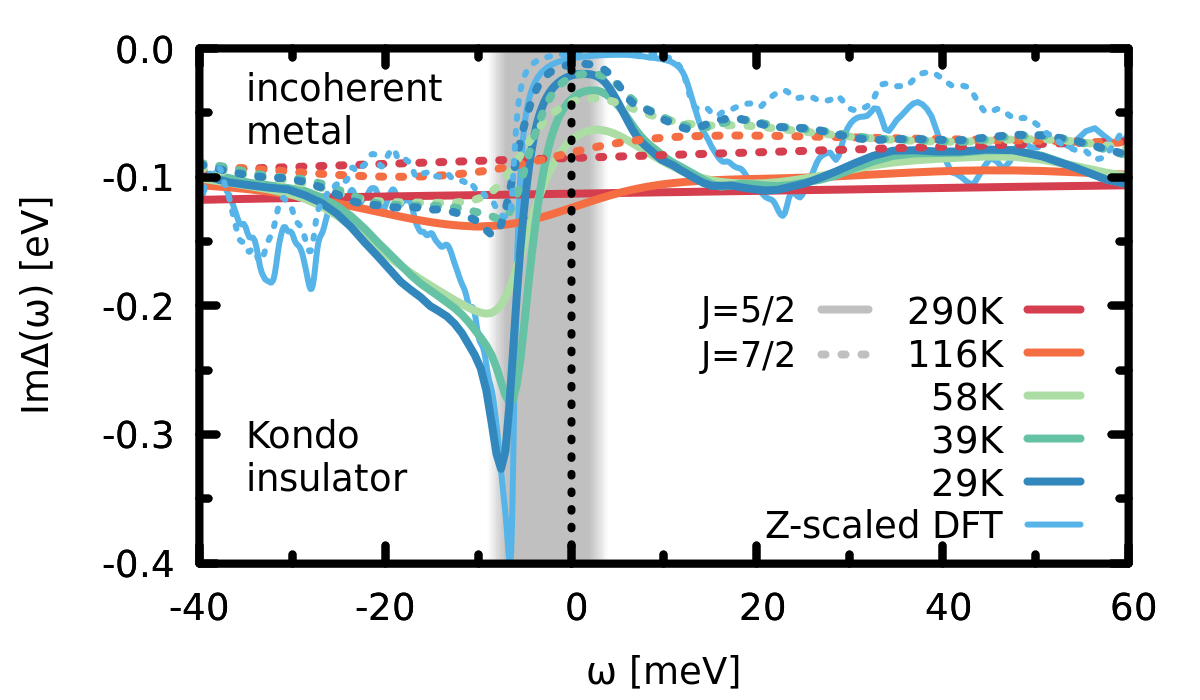}
\hspace{-0.4cm}
 \includegraphics[width=0.345\textwidth]{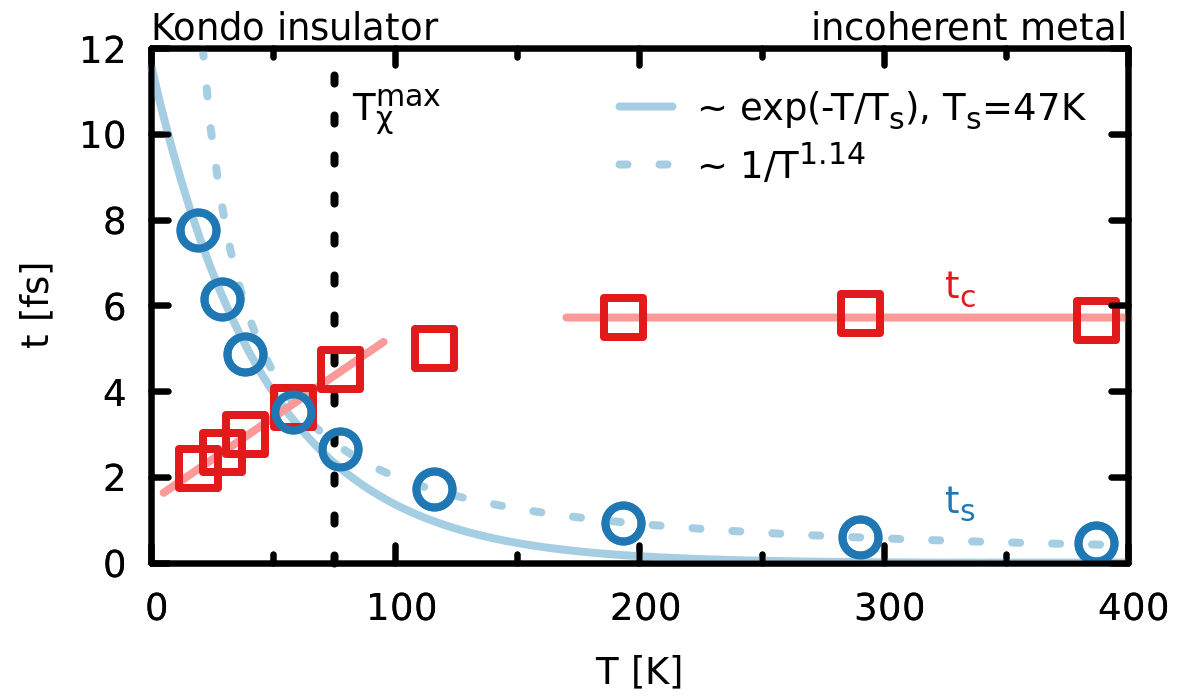}
\hspace{-0.25cm}
 \caption{(color online). 
Microscopic insights:
(a) Effective atom description: probability to find the system in a local state with $N$ particles and total angular momentum $J$. The histogram is virtually independent of temperature
(left columns: 29K, right columns: 290K). %
(b) Hybridization
function $\Delta(\omega)$: solid (dashed) lines represent the average over the $J=5/2$ ($J=7/2$) components. The shaded area delimits the charge gap (cf.~\fref{spec}(a)). For comparison the DFT hybridization functions $\Delta_0(Z\omega)$ are shown, with energies $\omega$ scaled with the quasi-particle weight $Z\approx 1/10$.
(c) Time-scales $t_s$ and $t_c$ of magnetic screening and valence fluctuations, respectively.
For both, we can distinguish two relaxation regimes separated by the common scale $T^{max}_\chi$.
}
 \label{micro}
\end{figure*}

By scrutinizing both $H^{atom}$ and $\Delta(\omega)$ we can gain insight into the fabric of the many-body state realized in \cbp%
\footnote{We only discuss the average of the $J=5/2$ states. For properties resolved into $m_J$, see Ref.~\cite{supps}.}.
First, we analyse the probability distribution of finding the solid in an eigenstate of $H^{atom}$, decomposed into 
local quantum numbers.
For example, we can distinguish atomic states according to their ($4f$) occupation $N$ and total angular momentum $J$. We find that \cbp\ is dominated 
by states
with $(N,J)=(1,5/2)$, see \fref{micro}(a):
The probabilities for said states sum to above 80\%.
Using the atomic states and their probabilities, we can also compute expectation values of local operators $\left\langle \mathcal{O}\right\rangle$:
We find that, at 29K, the impurity occupation is on
average $\left\langle N\right\rangle=1.02$ with significant deviations: $\delta N=\sqrt{\left\langle(N-\left\langle N\right\rangle)^2\right\rangle}=0.4$,
indicating that the $4f$-states in \cbp\ are far from being localized.
The average total angular momentum is $\left\langle J\right\rangle=2.52$, with a notable standard deviation $\delta J=0.8$ at 29K.
Similarly, we can compute instantaneous local correlators
$\left\langle\mathcal{O}^2\right\rangle:=\lim_{\tau\rightarrow 0}\left\langle \mathcal{O}(\tau)\mathcal{O}(0)\right\rangle$.
For example, we extract an instantaneous (disordered) moment
$\mu_{inst}=g_J\sqrt{\left\langle J^2\right\rangle}\mu_B/\hbar=g_J\sqrt{\left\langle j(j+1)\right\rangle}\mu_B\approx 2.65\mu_B$ (using $g_{J=5/2}=0.857$).
This moment is consistent with Curie-Weiss-fits, $\mu_{eff}^2/(3k_BT)$, to the experimental susceptibility\cite{PhysRevB.42.6842,NGCS} %for $T\gg T^{max}_\chi$ 
at high temperature (see \fref{arpessus}(c))
and very close to 
$\mu_{Ce^{3+}}=2.54\mu_B$, realized in isolated Ce$^{3+}$ ($J=5/2$) ions.
As seen in \fref{micro}(a), the probability histogram is virtually independent of temperature. As a consequence also the above instantaneous expectation values
hardly evolve when transitioning from the incoherent metal %($T>T^{max}_\chi$) 
to the Kondo insulating regime. % ($T<T^{max}_\chi$).
For example, the moment $\mu_{inst}$ increases by only 14\% when cooling from 290K to 29K%
\footnote{This situation is akin to the covalent insulator FeSi\cite{jmt_fesi}, but different from the ionic insulator LaCoO$_3$,
where a temperature-dependent histogram\cite{PhysRevB.86.195104} points to the occurrence of a spin-state transition.}.
The experimental uniform susceptibility $\chi(\svek{q}=0)$\cite{PhysRevB.42.6842} and the theoretical local magnetic susceptibility $\chi_{loc}$,
on the other hand, vary in the same temperature range by more than 50\%---a salient signature of a Kondo insulator\cite{NGCS}:
At high temperatures, see \fref{arpessus}(c), $\chi_{loc}$ and $\chi(\svek{q}=0)$
display Curie-Weiss behaviour. At intermediate temperatures, the susceptibilities peak at crossover temperatures $T^{max}_\chi$ and diminish upon cooling further%
\footnote{The low-$T$ upturn in $\chi(\svek{q}=0)$ is extrinsic\cite{PhysRevB.44.6832}}.
Quite generically, in the presence of a gap $\Delta$, $\chi(\vek{q}=0)$ vanishes in an activated fashion, while $\chi_{loc}(T=0)\propto \Delta^{-1}$ saturates.
That $\chi_{loc}>\chi(\svek{q}=0)$ for all temperatures  indicates the dominance of Hubbard interaction-driven
spin-fluctuations (see also Ref.\ \onlinecite{PhysRevB.78.033109}). The opposite is true in Hund's-physics controlled FeSi, where $\chi_{loc}\ll\chi(\svek{q}=0)$\cite{NGCS}.

How can we reconcile these seemingly disparate temperature evolutions: 
inert disordered moment $\mu_{inst}$ vs.\ strongly temperature dependent susceptibilities?
The answer lies in the time-scales of the spin-response: 
We have to go beyond the {\it instantaneous} correlators of $H^{atom}$, and allow for retardation effects induced by the hybridization $\Delta(\omega)$ to the particle bath.
Indeed, the static local susceptibility 
shown in \fref{arpessus}(c)
can be obtained by an (imaginary) time average: $\chi_{loc}(\omega=0)=\int_0^\beta d\tau\chi(\tau)\propto\int_0^\beta d\tau \left\langle J_z(\tau)J_z(0)\right\rangle$, where $\beta^{-1}=k_BT$.
The strong $T$-dependence in $\chi_{loc}(\omega=0)$ under the constraint of $\partial_T\mu_{inst}\approx0$, then points to
a strongly varying relaxation of local spin-fluctuations%
\footnote{See also Refs.~\cite{PhysRevLett.104.197002,PhysRevB.86.064411} and  Watzenb{\"o}ck \etal\ (in preparation) for the screening of local moments in iron pnictides.}.
We can provide systematic insight by deconstructing the local susceptibility\cite{PhysRevLett.115.247001,supps}:
\begin{equation}
\chi_{loc}(\omega=0)= \beta \chi_{loc}(\beta/2) + \int_0^\beta d\tau \left[\chi_{loc}(\tau)-\chi_{loc}(\beta/2)\right]
\label{chidecomp}
\end{equation}
as indicated by the shaded regions in \fref{arpessus}(c). The first term (green shaded part) describes the long-term behaviour, i.e.\ contributions that do not decay in time.
These low-energy excitations give rise (under a probing field) to Curie-Weiss behaviour with a saturated disordered local moment. 
In fact, in systems with Kondo screening, the temperature dependence in the uniform susceptibility is often analysed in terms of 
phenomenological Curie-Weiss-law with a $T$-dependent effective moment: $\chi(\vek{q}=0,T)=\mu_{eff}^2(T)/(3k_BT)$\cite{PhysRevLett.72.522,supps}.
The latter moment approaches
$\mu_{Ce^{3+}}$ in the high temperature single-ion limit as $J_z$ becomes a good quantum number (neglecting crystal-field effects):
The local $f$-moment is (asymptotically) unmoored from the conduction electrons' spin.
Below the coherence temperature $T^{max}_\chi$, however,  Kondo screening sets in: the moment decays,  approaching zero as spin excitations become gapped in the Kondo-insulating phase.
Interestingly, we here observe that 
the contribution $\beta\chi_{loc}(\beta/2)$ of the persisting local moment 
to the {\it local} susceptibility in fact
quantitatively mimics the experimental {\it uniform} susceptibility, see \fref{arpessus}(c). 
This finding suggests to associate the empirical effective moment, $\mu_{eff}(T)$, of the uniform susceptibility with the long-term memory of the local susceptibility: $\mu_{eff}(T)=\sqrt{3\chi_{loc}(\beta/2)}$.
We hence propose the computationally easier $\beta\chi_{loc}(\tau=\beta/2)$ as a proxy for $\chi(\svek{q}=0,\omega=0)$
in systems with Kondo screening.

The second contribution to $\chi_{loc}(\omega=0)$ in \eref{chidecomp} then describes the dynamical screening of the instantaneous moment 
to its magnitude persisting at large time-scales. 
At high temperatures, this term is small (see the grey shaded area in \fref{arpessus}(c)) as screening is poor and $\mu_{eff}$ is close to $\mu_{inst}$.
At lower temperatures, the screened moment $\mu_{eff}(T)$ 
decreases while $\mu_{inst}\approx const.$, as discussed above.
Thus the instantaneous moment %$\mu_{inst}$ 
is required to quench via efficient screening processes on short time-scales.
Fitting an exponential, $\chi(\tau)=\chi(\beta/2)+\left[\chi(0)-\chi(\beta/2)\right]\exp(-\tau/t_s)$, to the short-time decay, $\tau\le\beta/800$, 
we find two distinct relaxation regimes, see \fref{micro}(c): 
If Kondo screening is absent, i.e.\ in the incoherent metal regime above $T^{max}_\chi$, the characteristic time-scale $t_s$ follows hyperbolic growth with decreasing temperature.
When cooling below $T^{max}_\chi$, however, the rise in $t_s$ slows down and continues to grow in an activated fashion with a characteristic temperature
$T_s=47$K$=4$meV$/k_B$, corresponding to slightly less than half of the charge and spin gap. In the low temperature limit, the screening time reaches $t_s\approx 12$fs.
Indeed, in the Kondo insulating regime with spin-gap $\Delta_s\approx 12$meV\cite{PhysRevB.44.6832}, high-energy fluctuations are permissible 
for time-scales $t_s <\hbar/(2\Delta_s)\approx 27$fs (Heisenberg uncertainty principle). The magnitude of these fluctuations shoots up below $T^{max}_\chi$ and
 dominates the static local susceptibility at low enough temperatures, see \fref{arpessus}(c).
At absolute zero, this contribution is inversely proportional to the fundamental gap and, thus, the binding energy of the Kondo singlets.
While we cannot give a reliable value for $\chi_{loc}(T\rightarrow 0)$, we can estimate the Kondo temperature via the Kondo lattice expression
$T_K=4T^{max}_\chi/(2J+1)$\cite{PhysRevB.28.5255}. Using the maximum in the local-moment contribution $\beta \chi_{loc}(\beta/2)$ and $J=5/2$, we find $T_K\approx 50$K, in agreement with previous experimental estimates\cite{PhysRevB.42.6842}.

The Kondo scale can also be extracted from the charge degrees of freedom:
$k_BT_K=-\pi/4\Im\Delta_0\times Z/(2J+1)$
\cite{hewson,PhysRevB.28.5255}. Here $\Delta_0$ is the unrenormalized (DFT) hybridization function, and $Z\approx 1/10$ the quasi-particle weight.
Evaluating $\Delta_0$  at its peak-position at the edge of the gap (see \fref{micro}(b) and below), we find
$T_K=60$K---in excellent congruence with the estimate from the magnetic susceptibility.
The many-body hybridization function $\Delta(\omega)$ of \cbp\ itself displays a pronounced temperature dependence, see \fref{micro}(b):
At room temperature,  $\Im\Delta(\omega)$ is featureless---indicating  valence fluctuations in the local-moment regime to be incoherent.
In fact, the associated time-scale,  $t_c=\hbar/{\Im\Delta}$, remains constant at $\sim 6$fs throughout the incoherent metal regime, see \fref{micro}(c).
Upon cooling below $T^{max}_\chi$, an increasingly sharp minimum emerges in $\Im\Delta(\omega)$ at the edge of the gap. 
This enhancement signals the occurrence of the Kondo effect, engendering a coherent coupling between $4f-$ and conduction electron states.
As a consequence, the time-scale of valence fluctuations speeds up. We find: $t_c(T)-t_c(0)\propto T$.
In fact, at lowest temperatures, $\Im\Delta(w)$ approaches the bare (non-interacting) hybridization function $\Delta_0(Z\omega)$, when accounting for the mass renormalization by scaling its energy-dependence with $Z$: 
many-body excitations resemble a (strongly renormalized) band-structure, cf.\ \fref{arpessus}(b). 

The hybridization function thus plays a key role in the gap-formation.
In fact, on the basis of \eref{SCC}, we can distinguish three microscopic mechanisms for realizing an insulator, 
i.e.\ $\Im G_{loc}(\omega=0)=0$, without broken symmetries:
(i) Mott insulator: $H(\vek{k})$ is metallic, while $\Sigma(\omega)$ diverges inside the charge gap (Brinkmann-Rice scenario) and---as a consequence---$\Im\Delta(\omega)$ is suppressed; 
(ii) ionic band-insulator:
$H(\vek{k})$ is gapped by virtue of its crystal-fields $H^{loc}$, both $\Sigma$ and $\Delta$ are well-behaved at the Fermi level;
(iii) Kondo insulator: $H(\vek{k})$ is gapped, its crystal-fields $H^{loc}$ are irrelevant, $\Sigma$ is well-behaved but $\Delta(\omega)$
has a pronounced peak inside the gap to suppress coherent states.
That the typical behaviours of the  self-energy $\Sigma(\omega)$ and the hybridization function $\Delta(\omega)$ swap between a Mott and a Kondo insulator
has important consequences for the time-scale of fluctuations:
In their respective (incoherent) high-temperature regimes, a Mott and a Kondo insulator cannot be distinguished by looking at the magnitude of their
spin and valence relaxation times [for both: $t_c\sim 10t_s\sim\mathcal{O}(10\hbox{fs})$].
When cooling the Mott insulator, the freezing of charge fluctuations is heralded by an increase of $t_c$ by many orders of magnitude, 
while the relaxation of spins remains essentially unimpaired in the absence of a spin-gap. In this sense, spin and charge fluctuations are antagonistic, and for all $T$: $t_c\gg t_s$%
\cite{supps}.
In Kondo insulators, on the contrary, spin and charge degrees of freedom are indelibly interwoven by the Kondo effect:
For example, the transferred optical spectral weight (charge) neatly correlates with the effective local moment (spin)\cite{PhysRevLett.72.522,NGCS}.
Here, we demonstrate that this tie extends also to %the nature of %quiddity of
relaxation processes:
The distinct regimes of valence and spin fluctuations are both separated by the {\it common} scale $T^{max}_\chi$, close to which also the hierarchy
of time-scales switches ($t_s>t_c$ for $T\lesssim T^{max}_\chi$, $t_s<t_c$ for $T\gtrsim T^{max}_\chi$).
Importantly, we establish that the overall time-scale of valence fluctuations in \cbp, $t_c=1-6$fs,
is close to current time-resolution limits of ultrafast %core-level 
X-ray spectroscopies\cite{Hentschel2001,Buzzi2018}  as well as attosecond laser-pulse-based pump-probe measurements\cite{Goulielmakis2010}.
This suggests that monitoring the change in valence fluctuations through the Kondo crossover
and the sensitivity of Kondo singlet formation to external perturbations
 could soon be possible with time-resolved probes.

In conclusion, we computed spectral, optical and magnetic properties of \cbp\ from first principles and found excellent agreement with
experiment. We then extended the phenomenology of Kondo insulators by characterizing relaxation regimes of valence fluctuations and magnetic screening.
Delimiting the range of relevant time-scales, we motivated future applications of spectroscopies in the time domain,
which are the next step to elucidate the many-body fabric of Kondo insulators.

%%%%%%%%%%%%%%%%%%%%%%%%%%%%%%%%%%%%%%%%%%%%%%%%%%%%%%%%%%%%%%%%%%%%%%%%%%%%%%%%%%%%%%%%%%%%%%%%%%%%%%%%%%%%%%%%%%%%%%%%%%%%%%%%
%%%%%%%%%%%%%%%%%%%%%%%%%%%%%%%%%%%%%%%%%%%%%%%%%%%%%%%%%%%%%%%%%%%%%%%%%%%%%%%%%%%%%%%%%%%%%%%%%%%%%%%%%%%%%%%%%%%%%%%%%%%%%%%%
%%%%%%%%%%%%%%%%%%%%%%%%%%%%%%%%%%%%%%%%%%%%%%%%%%%%%%%%%%%%%%%%%%%%%%%%%%%%%%%%%%%%%%%%%%%%%%%%%%%%%%%%%%%%%%%%%%%%%%%%%%%%%%%%

 \paragraph{Acknowledgements.---}
\begin{acknowledgments}
The author gratefully acknowledges discussions with S.\ Dzsaber, G.\ Eguchi, K.\ Held, S.\ Paschen, F.\ Steglich, and A.\ Toschi. 
This work has been supported by the Austrian Science Fund (FWF)
through project ``LinReTraCe'' P~30213-N36.
\end{acknowledgments}

%\bibliography{../../refs,../../refs_mine}


\end{document}